\documentclass[12pt,reprint,amsmath,amssymb,aps,
,superscriptaddress,nofootinbib]{revtex4-1}

\usepackage{graphicx}
\usepackage{dcolumn}
\usepackage{bm}
\graphicspath{{./Figures/}}
\usepackage{graphicx}
\usepackage{float}
\usepackage{hyperref}
\usepackage{bm}
\usepackage{epstopdf}
\usepackage{overpic}
\usepackage{ulem}
\usepackage{color}
\usepackage{natbib} 
\usepackage{ulem} 

\usepackage[paperwidth=210mm,paperheight=297mm,hmargin=1.8cm,vmargin=2.61cm]{geometry}

\usepackage{titlesec}
\usepackage{etoolbox}
\makeatletter
\patchcmd{\ttlh@hang}{\parindent\z@}{\parindent\z@\leavevmode}{}{}
\patchcmd{\ttlh@hang}{\noindent}{}{}{}
\makeatother
\titlespacing{\section}{2pt}{\parskip}{\parskip}
\titlespacing{\subsection}{2pt}{\parskip}{\parskip}

\newcommand{\bff}{{\boldsymbol f}}

\newcommand{\OR}{\mathcal{O}}

\newcommand{\br}{\boldsymbol{r}}

\newcommand{\pll}{{\parallel}}

\begin{document}

\title{Morphological transitions of elastic filaments in shear flow
	}

\author{Yanan Liu}
\thanks{These two authors contributed equally.}
\affiliation{ESPCI Paris, PSL Research University, CNRS, Universit\'e Paris Diderot, Universit\'e Pierre et Marie Curie, Physique et M\'ecanique des Milieux H\'et\'erognes, UMR 7636, Paris, 75005, France.}
\author{Brato Chakrabarti}
\thanks{These two authors contributed equally.}
\affiliation{Department of Mechanical and Aerospace Engineering, University of California San Diego,
	9500 Gilman Drive, La Jolla, CA 92093, USA}
\author{David Saintillan}
\affiliation{Department of Mechanical and Aerospace Engineering, University of California San Diego,
	9500 Gilman Drive, La Jolla, CA 92093, USA}
\author{Anke Lindner}
\thanks{Corresponding author e-mail: anke.lindner@espci.fr}
\affiliation{ESPCI Paris, PSL Research University, CNRS, Universit\'e Paris Diderot, Universit\'e Pierre et Marie Curie, Physique et M\'ecanique des Milieux H\'et\'erognes, UMR 7636, Paris, 75005, France.}
\author{Olivia du Roure}
\thanks{Y.L., A.L. and O.dR. performed experiments. B.C. and D.S. performed simulations. All authors contributed to the theoretical model, to the analysis and interpretation of data, and to the manuscript preparation.}
\affiliation{ESPCI Paris, PSL Research University, CNRS, Universit\'e Paris Diderot, Universit\'e Pierre et Marie Curie, Physique et M\'ecanique des Milieux H\'et\'erognes, UMR 7636, Paris, 75005, France.}

\date{\today}

\begin{abstract}
\begin{center}
\textbf{ABSTRACT\\}
\end{center}
The {morphological} dynamics, instabilities and transitions of elastic filaments in viscous flows underlie a wealth of biophysical processes from flagellar propulsion to intracellular streaming, and are also key to deciphering the rheological behavior of many complex fluids and soft materials.\ Here, we combine experiments and computational modeling to elucidate the dynamical regimes and morphological transitions of elastic Brownian filaments in a simple shear flow.\ Actin filaments are employed as an experimental model system and their conformations are investigated through fluorescence microscopy in microfluidic channels.\ Simulations matching the experimental conditions are also performed using inextensible Euler-Bernoulli beam theory and non-local slender-body hydrodynamics in the presence of thermal fluctuations, and agree quantitatively with observations.\ We demonstrate that filament dynamics in this system is primarily governed by a dimensionless elasto-viscous number comparing viscous drag forces to elastic bending forces, with thermal fluctuations only playing a secondary role.\ While short and rigid filaments perform quasi-periodic tumbling motions, a buckling instability arises above a critical flow strength.\ A second transition to strongly-deformed shapes occurs at a yet larger value of the elasto-viscous number and is characterized by the appearance of localized high-curvature bends that propagate along the filaments in apparent ``snaking'' motions.\  A theoretical model for the so far unexplored onset of snaking  accurately predicts the transition and explains the observed dynamics. We present a complete characterization of filament morphologies and transitions as a function of elasto-viscous number and scaled persistence length and demonstrate excellent agreement between theory, experiments and simulations.

\end{abstract}
\maketitle

\section{Introduction}
The dynamics and conformational transitions of elastic filaments and semiflexible polymers in viscous fluids underlie the complex non-Newtonian behavior of their suspensions \cite{LS2015}, and also play a role in many small-scale biophysical processes from ciliary and flagellar propulsion \cite{Brennen77,Blake01} to intracellular streaming \cite{Ganguly12,Suzuki17}. The striking rheological properties of polymer solutions hinge on the microscopic dynamics of individual polymers, and particularly on their rotation, stretching and deformation under flow in the presence of thermal fluctuations. Examples of these dynamics include the coil-stretch \cite{de1974coil,schroeder2003observation} and stretch-coil \cite{young2007stretch,kantsler2012fluctuations} transitions in pure straining flows, and the quasi-periodic tumbling and stretching of elastic fibers and polymers in shear flows \cite{tornberg2004simulating,Schroeder05}. Elucidating the physics behind these microstructural instabilities and transitions is key to unraveling the mechanisms for their complex rheological behaviors \cite{bird1977dynamics}, from shear thinning and normal stress differences  \cite{becker2001instability} 
to viscoelastic instabilities \cite{Shaqfeh96} and turbulence~\cite{Morozov07}.

The case of long-chain polymers such as DNA \cite{schroeder2018single}, for which the persistence length $\ell_p$ is much smaller than the contour length $L$, has been characterized extensively in experiments \cite{perkins1997single,schroeder2003observation} as well as numerical simulations \cite{hur2000brownian} and mean-field models \cite{gerashchenko2006statistics}. The dynamics in this case is governed by the competition between thermal entropic forces favoring coiled configurations and viscous stresses that tend to stretch the polymer in strain-dominated flows. The interplay between these two effects is responsible for the coil-stretch transition in elongational flows and tumbling and stretching motions in shear flows, both of which are well captured by classic entropic bead-spring models \cite{Schroeder04,schroeder2005characteristic,Hsieh05}.

On the contrary, the dynamics of shorter polymers such as actin filaments \cite{harasim2013direct}, for which $L\sim\ell_p$, has been much less investigated  and is still not fully understood. Here, it is the subtle interplay of bending forces, thermal fluctuations and internal tension under viscous loading that instead dictates the dynamics. Indeed, bending energy and thermal fluctuations are now of comparable magnitudes, while the energy associated with stretching is typically much larger due to the small diameter of the {molecular} filaments \cite{LS2015}.\ This distinguishes these filaments from long entropy-dominated polymers such as DNA in which chain bending plays little role.

\begin{figure*}[t!]
	\centering
	\vspace{-0.2cm}
	\includegraphics[width=1\linewidth]{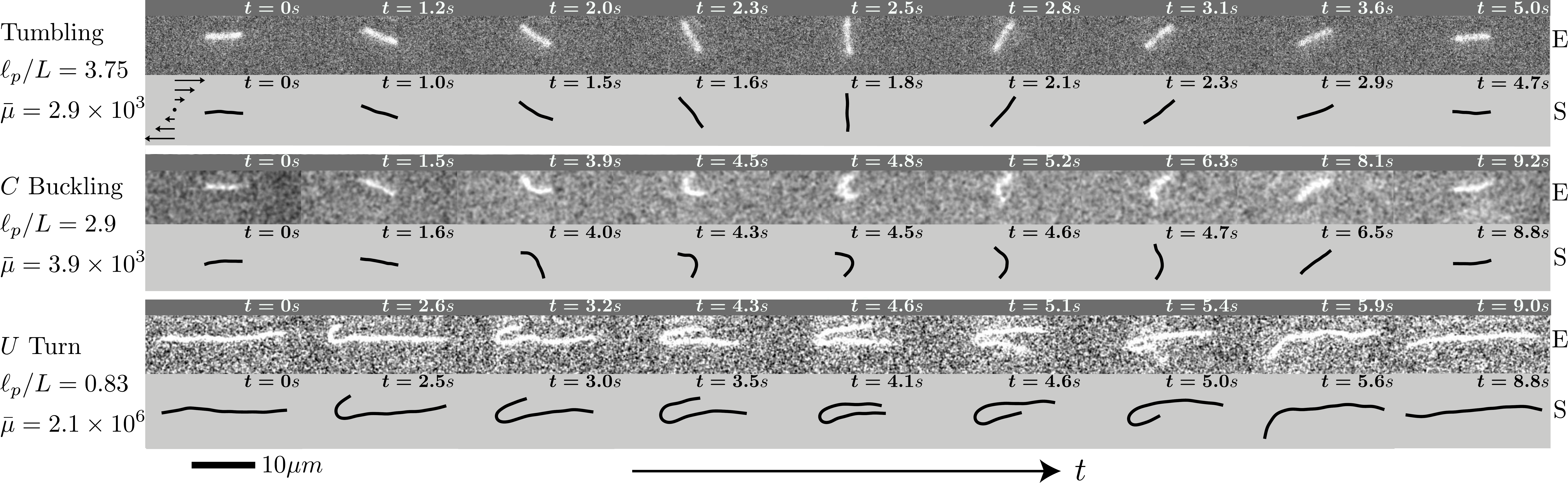}\vspace{-0.2cm}
	\caption{Temporal evolution of the filament shape in planar shear over one period of motion, showing three representative cases corresponding to increasing elasto-viscous numbers. In each case, we compare fluorescence images from experiments (E) to Brownian dynamics simulations (S). Movies of the dynamics are provided in the SI Appendix.} \vspace{-0.15cm}
	\label{Fig:snapshots}
\end{figure*}

The classical case of a rigid rod-like particle in a linear flow is well understood since the work of Jeffery \cite{jeffery1922motion}, who first described the periodic tumbling now known as Jeffery orbits occurring in shear flow. When flexibility becomes significant, viscous stresses applied on the filament can overcome bending resistance and lead to structural instabilities reminiscent of Euler buckling of elastic beams \cite{young2007stretch,tornberg2004simulating,manikantan2013subdiffusive,manikantan2015buckling,quennouz2015transport, kantsler2012fluctuations,Gugli12,manikantan2013subdiffusive}. 
On the other hand, Brownian orientational diffusion has been shown to control the characteristic period of tumbling \cite{harasim2013direct, lang2014dynamics}.
In  shear flow, the combination of rotation and deformation leads to particularly rich dynamics \cite{munk2006dynamics,harasim2013direct,nguyen2014hydrodynamics, forgacs1959particle,pawlowska2017lateral, delmotte2015, Chelakkot2010}, which have yet to be fully characterized and understood.

In this work, we elucidate these dynamics in a simple shear flow by combining numerical simulations, theoretical modeling and {model} experiments using actin filaments. The filaments we consider here have a contour length $L$ in the range of $4-40\,\mu$m and a diameter of $d\sim 8\,$nm. By analyzing the fluctuating shapes of the filaments, we measured the persistence length, as shown in \cite{gittes1993flexural}, to be $\ell_p = 17 \pm 1\, \mu$m independent of the solvent viscosity. We combine fluorescent labeling techniques, microfluidic flow devices and an automated-stage microscopy apparatus to systematically identify deformation modes and conformational transitions. Our experimental results are confronted against Brownian dynamics simulations and theoretical models that describe actin filaments as thermal inextensible Euler-Bernoulli beams whose hydrodynamics follow slender-body theory \cite{tornberg2004simulating}. By varying contour length as well as applied shear rates in the range of $\dot{\gamma} \sim 0.5-10\,$s$^{-1}$, we identify and characterize transitions from Jeffery-like tumbling dynamics of stiff filaments to buckled and finally strongly bent configurations for longer filaments. 

\section{Results and discussion}

\subsection{Governing parameters and filament dynamics}
In this problem, the filament dynamics results from the interplay of three physical effects -- {elastic bending forces}, thermal fluctuations and viscous stresses, and is governed by three independent dimensionless groups. {First, the ratio of the filament persistence length $\ell_p$ to the contour length $L$ characterizes the amplitude of transverse fluctuations due to thermal motion, with the limit of $\ell_p/L\rightarrow \infty$ describing rigid Brownian fibers}. Second, the elasto-viscous number $\bar{\mu}$ compares the characteristic time scale for elastic relaxation of a bending mode to the time scale of the imposed flow, and is defined in terms of the solvent viscosity $\mu$, applied shear rate $\dot{\gamma}$, filament length $L$ and bending rigidity $B$ as $\bar{\mu} = 8 \pi \mu \dot{\gamma} L^4 /B$. Note that $B$ and $\ell_p$ are related as $B=k_B T\ell_p$. Third, the anisotropic drag coefficients along the filament involve a geometric parameter $c = -\ln(\epsilon^2 \mathrm{e})$ capturing the effect of slenderness, where $\epsilon = d/L$. 

The elasto-viscous number can be viewed as a dimensionless measure of flow strength and exhibits a strong dependence on contour length. By varying $L$ and $\dot{\gamma}$, we have systematically explored filament dynamics over several decades of $\bar{\mu}$ and observed a variety of filament configurations, the most frequent of which we illustrate in Fig.~\ref{Fig:snapshots}. In relatively weak flows, the filaments are found to tumble without any significant deformation in a manner similar to rigid Brownian rods. On increasing the elasto-viscous number, a first transition is observed whereby compressive viscous forces overcome bending rigidity and drive a structural instability towards a characteristic $C$ shaped configuration during the tumbling motion. By analogy with Euler beams, we term this deformation mode {``global buckling''} as it occurs over the full length of the filament.
In stronger flows, this instability gives way to highly bent configurations, which we call $U$ turns and are akin to the snaking motions previously observed with flexible fibers \cite{forgacs1959particle,harasim2013direct}. During those turns, the filament remains roughly  aligned with the flow direction while a curvature wave initiates at one end and propagates towards the other end.\ At yet higher values of $\bar{\mu}$, more complex shapes can also emerge, including an $S$ turn which is similar to the $U$ turn but involves two opposing curvature waves emanating simultaneously from both ends (see SI Appendix for movies).\ In all cases, excellent agreement is observed between experimental measurements and Brownian dynamics simulations. Our focus here is in describing and explaining the first three deformation modes and corresponding transitions.

\begin{figure}[b!]
	\centering
	\vspace*{-0.45cm}
	\hspace*{-2mm}
	\includegraphics[scale=0.375]{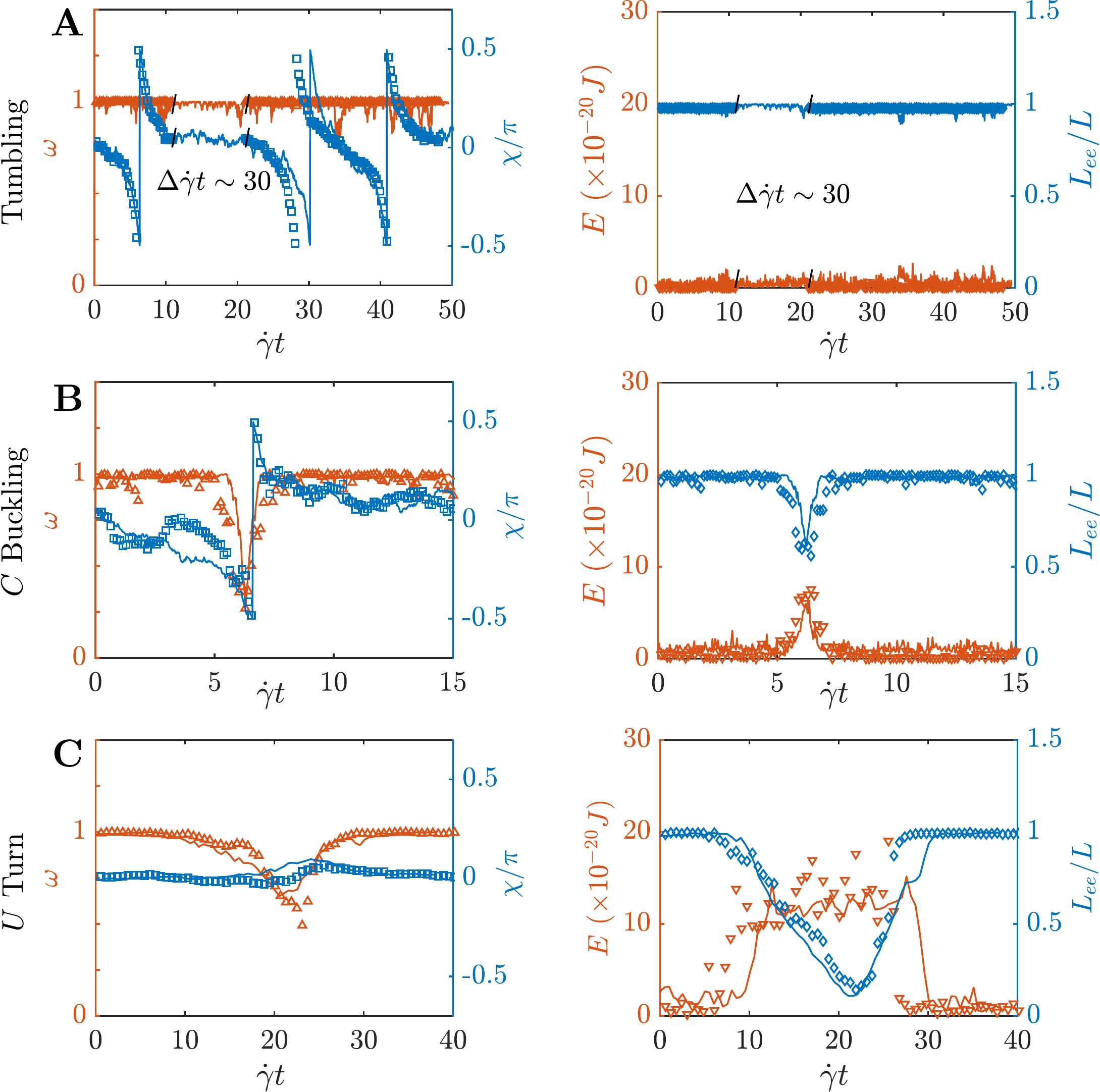}\vspace*{-0.1cm}
	\caption{Evolution of the sphericity parameter $\omega$, mean angle $\chi$ with respect to the flow direction, bending energy $E$ and scaled end-to-end distance $L_{ee}/L$ over one period of motion for (\textit{A}) Jeffery-like tumbling, (\textit{B}) $C$ buckling, and (\textit{C}) $U$ turn.  Symbols: experiments. Solid lines: simulations. Parameter values are the same as in Fig.~\ref{Fig:snapshots}. The lack of experimental data during the interval $\Delta \dot{\gamma}t\sim30$ in (\textit{A}) is due to a temporary loss of focus caused by tumbling of the filament out of the flow-gradient plane.}
	\label{Fig:COLB}
\end{figure}

\begin{figure*}[t!]
	\centering
	\includegraphics[width=0.99\linewidth]{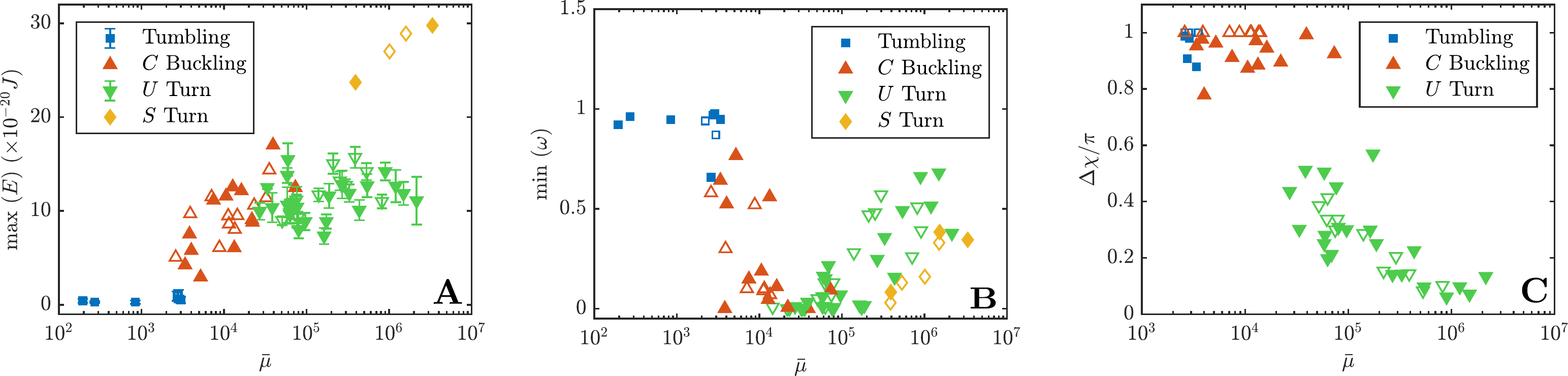}
	\vspace*{-0.25cm}
	\caption{Dependence on elasto-viscous number $\bar{\mu}$ of: (\textit{A}) the maximum value of the bending energy $E$, (\textit{B}) the minimum value of the sphericity parameter $\omega$, and (\textit{C}) the range $\Delta\chi$ of the mean angle in the various tumbling and deformation regimes. Full symbols: experiments; open symbols: simulations. For experimental data, the measurement error in $\bar{\mu}$ (due to errors in contour length ($\pm 0.5\,\mu$m) and in local shear rate  ($\pm 0.1\,$s$^{-1}$)) is comparable to the marker size. \vspace{-0.3cm}}
	\label{Fig:Evolution}
\end{figure*}

We characterize the temporal shape evolution more quantitatively for each case in Fig.~\ref{Fig:COLB}.\ In order to describe the overall shape and orientation of the filament, we introduce the gyration tensor, or the second {mass}  moment, as \vspace{-0.15cm}
\begin{equation}
{G}_{ij}(t)=\frac{1}{L}\int_{0}^{L}[r_i(s,t)-\bar{r}_i(t)][r_j(s,t)-\bar{r}_j(t)]\,ds, \vspace{-0.15cm}
\end{equation}
where $\boldsymbol{r}(s,t)$ is a two-dimensional parametric representation of the filament centerline with arclength $s\in[0,L]$ in the flow-gradient plane, and $\bar{\boldsymbol{r}}(t)$ is the instantaneous center-of-mass position.
{The angle $\chi$ between the mean filament orientation and the flow direction is provided by the eigenvectors of $G_{ij}$, while its eigenvalues $(\lambda_1,\lambda_2)$ can be combined to define a sphericity parameter $\omega=1 - 4 \lambda_1\lambda_2 /(\lambda_1+\lambda_2)^2$ quantifying filament anisotropy: $\omega\approx 0$ for nearly isotropic configurations ($\lambda_1\approx \lambda_2$), and $\omega\approx 1$ for nearly straight shapes ($\lambda_1\gg\lambda_2\approx 0$). Other relevant measures of filament conformation are the scaled end-to-end distance $L_{ee}(t)/L=|\boldsymbol{r}(L,t)-\boldsymbol{r}(0,t)|/L$, whose departures from its maximum value of 1 are indicative of bent or folded shapes, and the total bending energy $\smash{E(t) = \frac{B}{2}\int_0^L \kappa^2(s,t) ds}$, which is an integrated measure of the filament curvature $\kappa(s,t)$.}

As is evident in Fig.~\ref{Fig:COLB}, these different variables exhibit distinctive signatures in each of the three regimes {and can be used to systematically differentiate between configurations}.  During Jeffery-like tumbling, filaments remain nearly straight with $\omega\approx 1$, $L_{ee}\approx L$ and $E\approx 0$ while the angle $\chi$ quasi-periodically varies from $-\pi/2$ to $\pi/2$ over the course of each tumble.  During a $C$ buckling event, the angle $\chi$ still reaches $\pi/2$, but the other quantities now deviate from their baseline as the filament bends and straightens again.\ This {provides} a quantitative {measure} for distinguishing tumbling motion and $C$ buckling. {During a $U$ turn, however, deformations are also significant but $\chi$ only weakly deviates from $0$ as the filament remains roughly aligned with the flow direction and executes a tank-treading motion rather than an actual tumble. This feature provides a simple test for distinguishing $C$ and $U$ turns in both experiments and simulations.}\ Other hallmarks of $U$ turns are the increased bending energy during the turn, which exhibits a nearly constant plateau while the localized bend in the filament shape travels from one end to the other, and a strong minimum in the end-to-end distance $L_{ee}(t)$, which reaches nearly zero halfway through the turn when the filament is symmetrically folded.

\subsection{Order parameters}

This descriptive understanding of the dynamics allows us to investigate transitions between deformation regimes as the elasto-viscous number increases.
The dependence on $\bar{\mu}$ of the maximum bending energy $E$, minimum value of the sphericity parameter $\omega$, and range $\Delta\chi$ of the mean angle over one or several periods of motion is shown in Fig.~\ref{Fig:Evolution}. In the case of $U$ turns, the maximum bending energy is calculated as an average over the plateau seen in Fig.~\ref{Fig:COLB}\textit{C}. In the tumbling regime, deformations are negligible beyond those induced by thermal fluctuations, as evidenced by the nearly constant values of $\mathrm{max}(E)\approx 0$ and $\mathrm{min}(\omega)\approx 1$. After the onset of buckling, however, the maximum bending energy starts increasing monotonically with $\bar{\mu}$ as viscous stresses cause increasingly stronger bending of the filament.\ This increased bending is accompanied by a decrease in $\omega$ as bending renders shapes increasingly isotropic, finally reaching $\mathrm{min}(\omega)\approx 0$.\ Interestingly, the transition to $U$ turns is marked by a plateau of the bending energy, which subsequently only very weakly increases with $\bar{\mu}$.\ This plateau is indicative of the emergence of strongly bent configurations where the elastic energy becomes localized in one sharp fold, and suggests that the curvature of the folds during $U$ turns depends only weakly on flow strength. The parameter $\omega$ also starts increasing again after the onset of $U$ turns, as the filaments adopt hairpin shapes that become increasingly anisotropic. Figure~\ref{Fig:Evolution}\textit{AB} also shows a few data points for $S$ turns at high values of $\bar{\mu}$: in this regime, the maximum bending energy is approximately twice that of $U$ turns, as bending deformations now become localized in two sharp folds instead of one.\ $S$~shapes are, however, more compact than $U$ shapes and thus show lower values of $\omega$.

Orientational dynamics are summarized in Fig.~\ref{Fig:Evolution}\textit{C}, showing the range $\Delta \chi = \chi_\mathrm{max}-\chi_\mathrm{min}$ of the mean angle $\chi$ over one period of motion. During a typical Jeffery-like tumbling or $C$ buckling event, the 
main filament orientation rotates continuously and as a result $\Delta \chi = \pi$. The scatter in the experimental data  is the result of the finite sampling rate during imaging. During $U$ turns,  the filament no longer performs tumbles but instead remains globally aligned with the flow direction as it undergoes its snaking motion, resulting in $\Delta \chi < \pi$.
This explains the discontinuity in the data of Fig.~\ref{Fig:Evolution}\textit{C}, where $C$ and $U$ turns stand apart. 
As $\bar{\mu}$ increases  beyond the transition, 
we find that $\Delta \chi \to 0$ suggesting a nearly constant mean orientation for the folded shapes characteristic of $U$ turns.  

While we have not studied the tumbling frequency extensively, data based on a limited number of simulations and experiments recovers the classical 2/3 scaling of frequency on flow strength \cite{schroeder2005characteristic,harasim2013direct} for the explored range of parameters, with a systematic deviation towards 3/4 in strong flows in agreement with results from Lang et al.\ \cite{lang2014dynamics}.

\subsection{{Transitions between regimes and phase diagram}}

\begin{figure}[t!]
	\centering
	\includegraphics[scale=0.74]{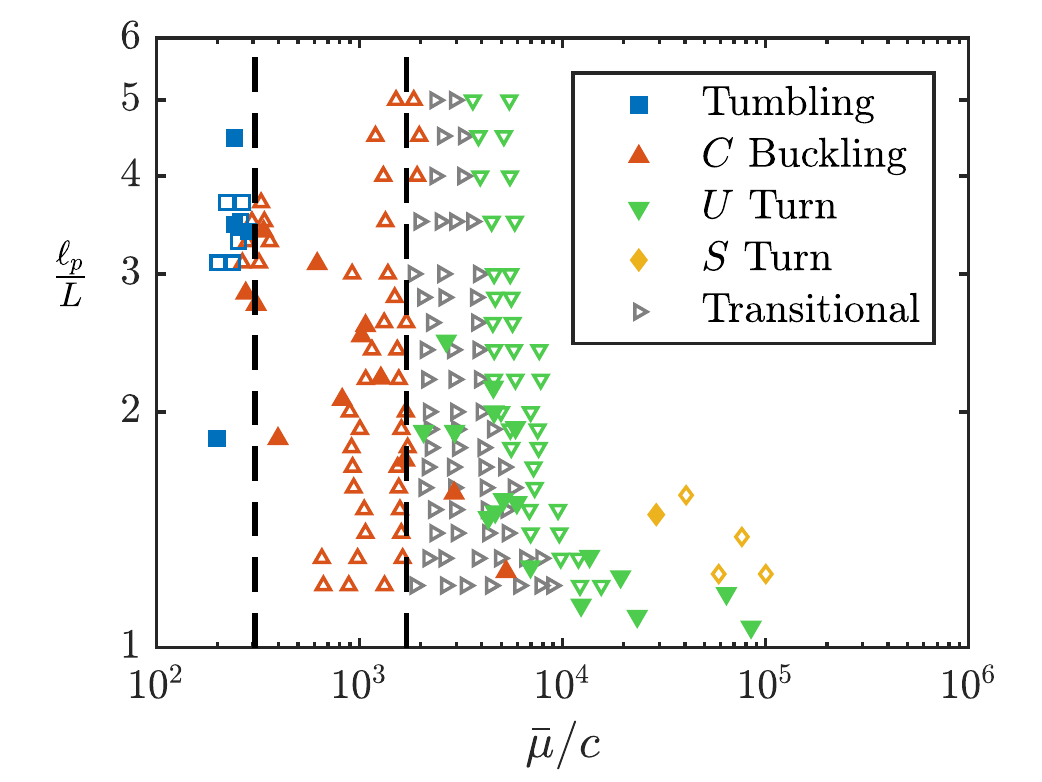}
	\vspace*{-3mm}
	\caption{{Phase chart indicating the different dynamical regimes in the $(\bar{\mu}/c,\ell_p/L)$ parameter space. The dashed black lines show the theoretical transitions from tumbling motion to  $C$ buckling ($\bar{\mu}_c^{(1)}/c \approx 306.4$), and from $C$ buckling to $U$ turns ($\bar{\mu}_c^{(2)}/c \approx 1700$). Full symbols: experiments; open symbols: simulations.}}\vspace{-0.25cm}
	\label{Fig:Phases}
\end{figure}

Our experiments and simulations have uncovered three dynamical regimes with increasing values of $\bar{\mu}$, the transitions between which we now proceed to explain. A summary of our results is provided in Fig.~\ref{Fig:Phases} as a phase diagram in the $(\bar{\mu}/c,\ell_p/L)$ parameter space, where the transitions are found to occur at fixed values of $\bar{\mu}/c$ independent of $\ell_p/L$. The first transition from tumbling motion to $C$ buckling has received much attention in the past, primarily in the case of non-Brownian filaments  \cite{becker2001instability, nguyen2014hydrodynamics, tornberg2004simulating}.\ This limit is amenable to a linear stability analysis \cite{becker2001instability}, which predicts a supercritical pitchfork bifurcation whereby compressive viscous stresses exerted along the filament as it rotates into the compressional quadrant of the flow are sufficiently strong to induce buckling. 
The stability analysis is based on local slender-body theory, where the natural control parameter arises as $\bar{\mu}/c$, and predicts buckling above a critical value of $\bar{\mu}_c^{(1)}/c \approx 306.4$ \cite{becker2001instability}, in reasonable agreement with our measurements (Fig.~\ref{Fig:Phases}).

Thermal fluctuations do not significantly alter this threshold, but instead result in a blurred transition \cite{baczynski2007stretching, kantsler2012fluctuations,manikantan2015buckling} with an increasingly broad transitional regime where both tumbling and $C$ buckling can be observed for a given value of $\bar{\mu}$. When Brownian fluctuations are strong, i.e., for low values of $\ell_p/L$, it becomes challenging to differentiate deformations caused by viscous buckling vs fluctuations, and thus the distinction between the two regimes becomes irrelevant.

\begin{figure}[b!]
	\centering
	\vspace*{-2mm}
	\includegraphics[width=8cm]{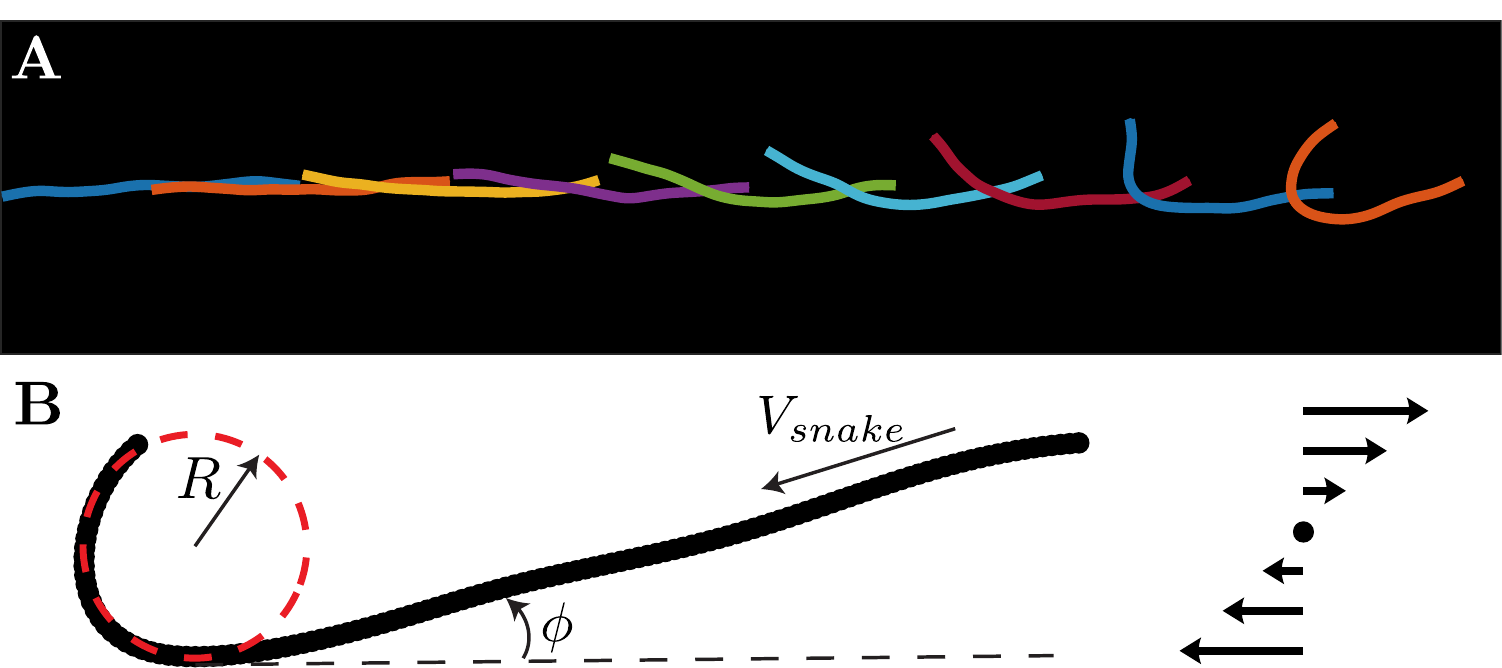}
	\caption{(\textit{A}) Numerical snapshots of filament shapes during the formation of a $J$ shape before the initiation of a $U$ turn. (\textit{B}) The $J$ shape can be approximated by a semicircle of radius $R$ connected to a straight arm forming a tilt angle of $\phi$ with the flow direction. During snaking, the filament translates tangentially with an axial velocity $V_{snake}$.}
	\label{Fig:J_shape}
\end{figure}

Upon increasing $\bar{\mu}/c$,  the second conformational transition from $C$ shaped filaments to elongated hairpin-like $U$ turns undergoing snaking motions occurs. The appearance of $U$ turns (shown in green in Fig.~\ref{Fig:Phases}) occurs above a critical value $\smash{\bar{\mu}^{(2)}_c/c}$ that is again largely independent of $\ell_p/L$.  However, the transition is not sharp, and near the critical value both shapes can be observed simultaneously (as indicated by gray points). In fact, a single filament in the transitional regime will typically execute both types of turns, switching stochastically between them (see SI Appendix, Fig. S6). This transition towards snaking dynamics has not previously been characterized.\ Our attempt at understanding its mechanism focuses on the onset of a $U$ turn, which always involves the formation of a $J$ shaped configuration as visible in Fig.~\ref{Fig:snapshots} and also illustrated in Fig.~\ref{Fig:J_shape}. 

To elucidate the transition mechanism, we develop a theoretical model for a $J$ configuration, which can be viewed as a precursor to the $U$ turn. We neglect Brownian fluctuations and idealize the $J$ shape as a semi-circle of radius $R$ connected to a straight arm forming an angle $\phi$ with the flow direction, with both sections undergoing a snaking motion responsible for the $U$ turn; details of the model, which draws on analogies with the tank-treading motion of vesicles \cite{keller_skalak_1982,rioual2004analytical}, can be found in the SI Appendix. By satisfying filament inextensibility as well as force and torque balances, and by balancing viscous dissipation in the fluid with the work of elastic forces, we are able to solve for model parameters such as $R$ and $\phi$ {without any fitting}.\ A key aspect of the model is that consistent solutions for these parameters can only be obtained above a critical elasto-viscous number, and this solvability criterion thus provides a threshold $\bar{\mu}_{c}^{(2)}/c \approx 1700$ below which the $J$ shape ceases to exist. This theoretical prediction is depicted by the dashed line in the phase chart of Fig.~\ref{Fig:Phases} and coincides perfectly with the onset of the transitional regime in simulations and experiments.

\begin{figure*}[t!]
	\centering
	\includegraphics[width=0.99\linewidth]{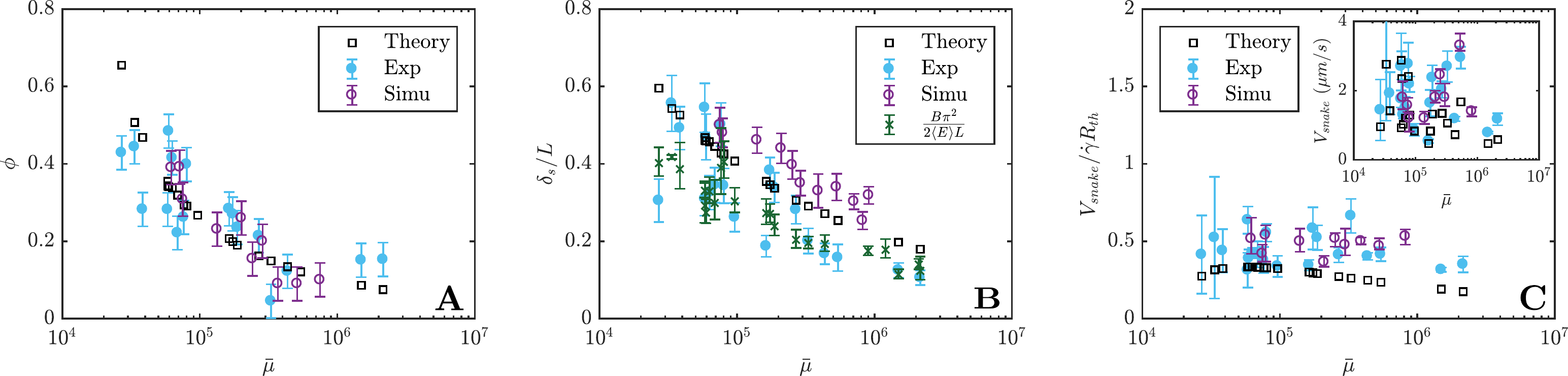}
	\vspace*{-0.3cm}
	\caption{(\textit{A}) Dependence on $\bar{\mu}$ of the tilt angle $\phi$ formed by $J$ shapes with respect to the flow direction in experiments, simulations and in our theoretical model. ($\textit{B}$) Fraction $\delta_s/L$ of the filament length that is bent during a $U$ turn (see Fig.~S4 for the detailed definition of $\delta_s$). The theoretical predictions are based on the $J$ shape at the start of the turn. Since the bending energy during a $U$ turn is concentrated in the fold, an estimate for $\delta_s/L$ is also provided by $B\pi^2/2\langle E\rangle L$ where $\langle E\rangle$ is the average bending energy during the turn, and good agreement is found between both measures. (\textit{C}) Snaking velocity $V_{snake}$ rescaled with $\dot{\gamma} R_{th}$ and plotted against $\bar{\mu}$ from experiments, simulations and theory; here, $R_{th}$ is the theoretically predicted fold radius.}
	\label{Fig:Uturn}\vspace{-0.25cm}
\end{figure*}

{We can now discuss the initiation of the $J$-shape, in which} two possible mechanisms may be at play. On the one hand, it may be caused by the global buckling of the filament in the presence of highly compressive viscous forces, in a manner consistent with the sequence of shapes of Fig.~\ref{Fig:J_shape}\textit{A}.
{Under sufficiently strong shear, compressive forces can induce a buckling instability on a filament that has not yet aligned with the compressional axis and forms only a small angle with the flow direction.} Alignment of the deformed filament with the flow then results in differential tension (compression vs tension) near its two ends, thus allowing one end to bend while the other remains straight.\ A second potential mechanism proposed in \cite{lang2014dynamics} is of a local buckling occurring on the typical length scale of transverse thermal fluctuations. Our data, however, clearly show that the transition to $U$ turns is independent of thermal fluctuations, allowing us to discard this {hypothesis}. Thermal fluctuations are nonetheless responsible for the existence of the transitional regime above $\bar{\mu}_{c}^{(2)}/c$, where they can destabilize $J$ shapes towards $C$ shapes and thus prevent the occurrence of $U$ turns. This interpretation is consistent with the increasing extent of the transitional regime with decreasing $\ell_p/L$.

\subsection{Dynamics of \textit{U}-turns}

We further characterize the dynamics during $U$ turns, for which our theoretical model also provides predictions. The filament orientation at the onset of a turn is plotted in Fig.~\ref{Fig:Uturn}\textit{A}, showing the tilt angle $\phi$ formed by the straight arm of the $J$ shape with respect to the flow direction as a function of $\bar{\mu}$. Our theoretical model for dynamics of the $J$ shape also provides the value of $\phi$, in excellent agreement with experiments. In both cases, the tilt angle decreases with increasing flow strength due to increased alignment by the flow. For very long filaments (limit of large $\bar{\mu}$), accurate measurements of the tilt angle become challenging due to shape fluctuations, hence the increased scatter in the data.

After a $J$ shape is initiated as discussed above, the curvature of the folded region remains nearly constant in time as suggested by the plateau in the bending energy (Fig.~\ref{Fig:Evolution}\textit{C}). This provides a strong basis for approximating the bent part of the filament as a semi-circle of radius $R$ in our model. The theoretical prediction $R_{th}$ and  measurements of the radius on $J$ shapes from experiments and simulations agree quite well in~Fig.~\ref{Fig:Uturn}\textit{B} (see SI Appendix for details). The radius of the bend is seen to decrease with $\bar{\mu}$, as compressive viscous stresses in strong flows allow increasingly tighter folding of the filament.

The rotation of the end-to-end vector during the $U$ turn results primarily from tank-treading of the filament along its arclength, unlike the global rotation that dominates the tumbling and $C$ buckling regimes.\ While the snaking velocity is not constant during a turn, its average value can be quantitatively measured through the time derivative of the end-to-end distance, yielding the approximation $V_{snake}\approx \dot{L}_{ee}/2$. The relevant dynamic length and time scales during this snaking motion are  the radius of curvature $R$ of the bent segment and shear rate $\dot{\gamma}$. This is  supported by our theory, where rescaling $V_{snake}$ by $\dot{\gamma}R_{th}$ collapses the predicted velocities over a range of filament lengths (SI Appendix, Fig. S3). The same rescaling applied to the experimental and numerical data and using the theoretical radius $R_{th}$ also provides a good collapse in Fig.~\ref{Fig:Uturn}\textit{C}.

Harasim \textit{et al.} \cite{harasim2013direct} previously proposed a simplified theory of the $U$ turn, which shares similarities with ours but assumes that the filament is aligned with the flow direction and neglects elastic stresses inside the fold. Their predictions are in partial agreement with our results in the limit of very long filaments and strong shear (see SI Appendix). Their theory is unable to predict {and explain} the transition from buckling to $U$ turns.


\section{Concluding remarks}

Using stabilized actin filaments as a model polymer, we have systematically studied and analyzed the conformational transitions of elastic Brownian filaments in simple shear flow as the elasto-viscous number is increased. Our experimental measurements were shown to be in excellent agreement with a computational model describing the filaments as fluctuating elastic rods with slender-body hydrodynamics. By varying filament contour length and applied shear rate, we performed a broad exploration of the parameter space and confirmed the existence of a sequence of transitions, from rod-like tumbling to elastic buckling to snaking motions. While snaking motions had been previously observed in a number of experimental configurations, the existence of a $C$ buckling regime had not been confirmed clearly. This is due to the fact that $C$ buckling is only visible over a limited range of elasto-viscous numbers and occurs only in simple shear flow, challenging to realize experimentally. We showed that both transitions are primarily governed by $\bar{\mu}/c$. Brownian fluctuations do not modify the thresholds but tend to blur the transitions by allowing distinct dynamics to coexist over certain ranges of $\bar{\mu}$.

While the first transition from tumbling to buckling had been previously described as a supercritical linear buckling instability \cite{becker2001instability}, the transition from buckling to snaking was heretofore unexplained. Using a simple analytical model for the dynamics of the $J$ shape that is the precursor to snaking turns, we were able to obtain a theoretical prediction for the threshold elasto-viscous number above which snake turns become possible. The model did not take thermal noise into account, but highlighted the subtle role played by tension and compression during the onset of the turn. Our analysis and model lay the groundwork for illuminating a wide range of other complex phenomena in polymer solutions, from their rheological response in flow and {dynamics in semi-dilute solutions \cite{Kirchenbuechler2014, Huber2014} to} migration under confinement and microfluidic control of filament dynamics.

\begin{acknowledgments}
We are grateful to Guillaume Romet-Lemonne and Antoine J\'egou for providing purified actin and to Thierry Darnige for help with the programming of the microscope stage. We  thank Michael Shelley, Lisa Fauci, Julien Deschamps, Andreas Bausch, Gwenn Boedec, Anupam Pandey, Harishankar Manikantan and Lailai Zhu for useful discussions, and Roberto Alonso-Matilla for checking some of our calculations. The authors acknowledge support from ERC Consolidator Grant No.\ 682367, from a CSC Scholarship, and from NSF Grant CBET-1532652. 
\end{acknowledgments}

\appendix

\section*{Appendix}

\section{Experimental Methods} 

The protocol
for assembly of the actin filaments  is well controlled and reproducible.
Concentrated G-actin, which is obtained from rabbit muscle and purified
according to the protocol described in \cite{spudich1971regulation},
is placed into F-Buffer (10$\,$mM Tris-Hcl pH$=$7.8, 0.2$\,$mM ATP,
0.2$\,$mM CaCl$_2$, 1$\,$mM DTT, 1$\,$mM MgCl$_2$, 100$\,$mM KCl,
0.2$\,$mM EGTA and 0.145$\,$mM DABCO) at a final concentration of
1$\,\mu$M. At the same time, Alexa488-fluorescent phalloidin in the same
molarity as G-actin is added to prevent depolymerization and thus to stabilize as well as to visualize the filaments.
After 1 hour of  polymerization in the dark at room temperature,
concentrated F-actin is stored at $4^{\circ}$C  for following
experiments. To avoid interactions between filaments, F-actin used in
experiments has a final concentration of 0.1$\,$nM obtained by diluting
the previous solution with F-buffer. 1$\,$mM ascorbic acid is added to
decrease photo-bleaching effects and 45.5\%(w/v) sucrose to match the
refractive index of the PDMS channel ($n=1.41$). The viscosity of the
dilute filament suspension is $5.6\,$mPa$\cdot$s at $20^{\circ}$C,
measured on an Anton Paar MCR 501 rheometer.

A micro PDMS channel is designed as a vertical Hele-Shaw cell, with 
length $L=30\,$mm, height $H=500\,\mu$m and width $W=150\,\mu$m. 
In this geometry the filament dynamics can be directly observed in the horizontal shear plane whereas shear in the vertical direction can be neglected at a sufficient distance from the bottom wall (see SI Appendix for more details). To consider  pure shear flow, filament and flow scales should be properly separated, and we thus chose a width (150 $\mu m$) much larger
than the typical dimension of the deformed filament ($\approx 10 \mu m$). An objective with long working-distance is required to observe in a plane far enough from the bottom; the objective  should also have a large numerical aperture to collect as much  light as possible from the fluorescent actin filaments. To combine both of these requirements, we have used a water immersion objective from Zeiss (63X C-Apochromat /1.2NA)
with WD$\,\approx280\,\mu$m. 

Stable flow is driven by a syringe pump (Cellix ExiGo) and 
particle tracking velocimetry has been used to check the agreement of the velocity profile with theoretical predictions \cite{white2006viscous}. We impose
flow rates $Q$ in the range of  $5-7.5\,$nL/s, leading to typical filament velocities $u_x\sim20-150\,\mu$m/s in the observation area in the
plane $z=150\,\mu$m.\ The filament Reynolds number is of the order $\mathrm{Re}\sim10^{-4}$. To follow the
filaments during their transport in the channel we use a motorized stage
programmed to accurately follow the flow and also to correct for small changes in the $z$ plane, occuring
due to slight bending of the channel. This step is necessary as the focal
depth of the objective is only of a few microns and streamlines need to be followed with high precision over distances of several mm.

Images are captured by a s-CMOS camera (HAMAMATSU ORCA flash 4.0LT, 16
bits) with an exposure time of $\Delta t=65\,$ms and are synchronized with
the stage displacement. They are processed by Image J to obtain the
position of the center of mass and the filament shape. The center of mass
is used to calculate the local shear rate experienced by the filament. The
shape is extracted through Gaussian blur, threshold, noise reduction and
skeletonize procedures. A custom MATLAB code is then used to reconstruct
the filament centerline as a sequence of discrete points along the arclength $s$ and to calculate the parameters plotted in Fig.~\ref{Fig:COLB}.

 \section{Modeling and Simulations}
We model the filaments as inextensible Euler-Bernoulli beams and use non-local slender-body hydrodynamics to capture drag anisotropy and hydrodynamic interactions \cite{tornberg2004simulating,manikantan2013subdiffusive}.\ Simulations without hydrodynamic interactions (free-draining model) were also performed but did not compare well with experiments. Brownian fluctuations are included and satisfy the fluctuation-dissipation theorem. As experiments only consider quasi-2D trajectories involving dynamics in the focal plane, we perform all simulations in 2D and indeed found better agreement compared to 3D simulations. Details of the governing equations and numerical methods  are provided in the SI Appendix. The simulation code is available upon request to the authors.


\normalem

\widetext
\clearpage
\pagebreak
\newpage

\begin{center}
	\textbf{\Large Supporting Information}
\end{center}
\setcounter{section}{0}
\setcounter{equation}{0}
\setcounter{figure}{0}
\setcounter{table}{0}
\makeatletter
\renewcommand{\theequation}{S\arabic{equation}}
\renewcommand{\thefigure}{S\arabic{figure}}

\section{Microfluidic channel geometry}
\begin{figure}[H]
	\centering
	\hspace*{0cm}
	\vspace*{-0.3cm}
	\begin{overpic}[width=7.7cm]{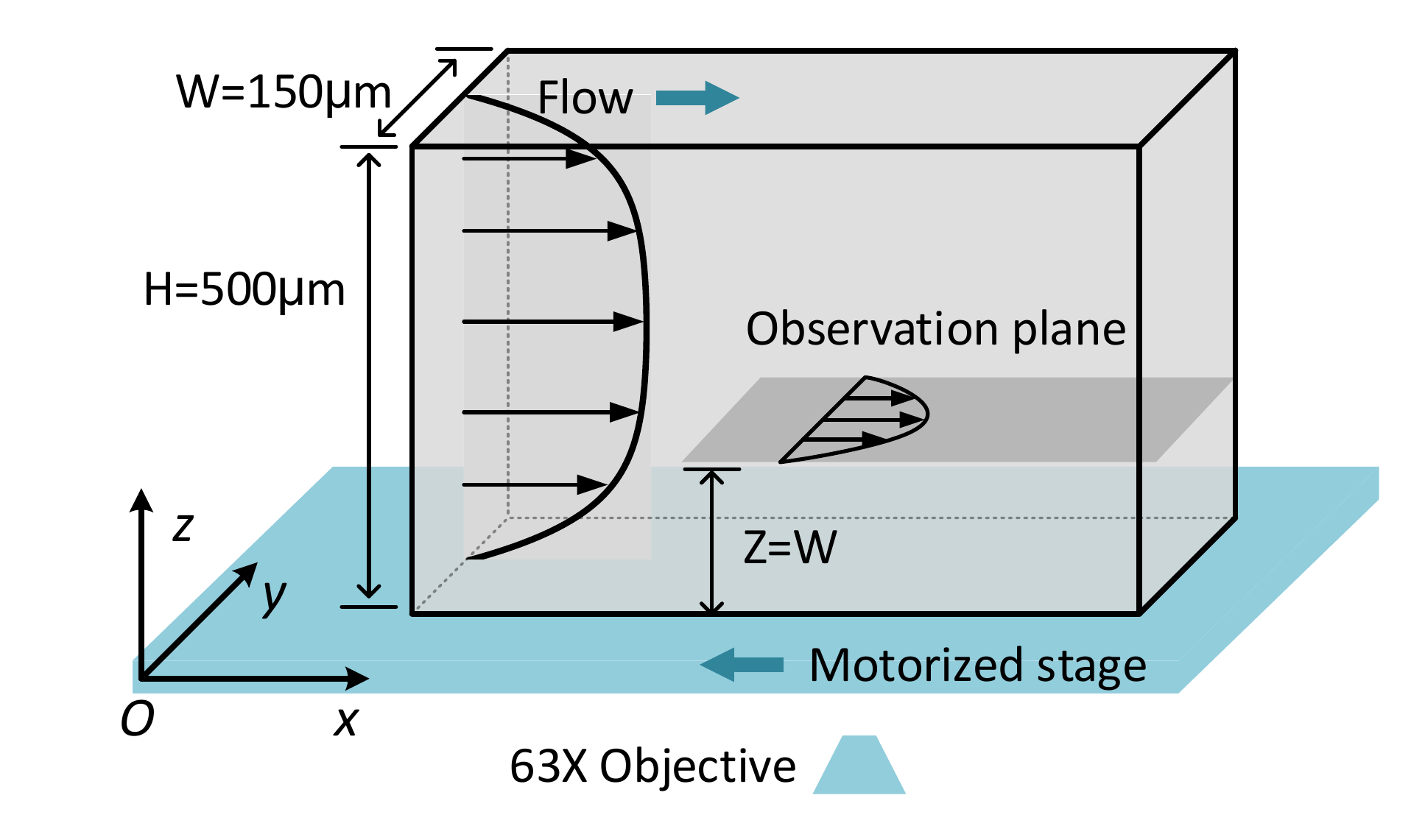}
		\put(0,52){\textsf{{A}}}
		\put(0,-5){\textsf{{B}}}
		\put(0,-37){\textsf{{C}}}
	\end{overpic}
	\vspace*{-0.3cm}
	\includegraphics[width=4.65cm]{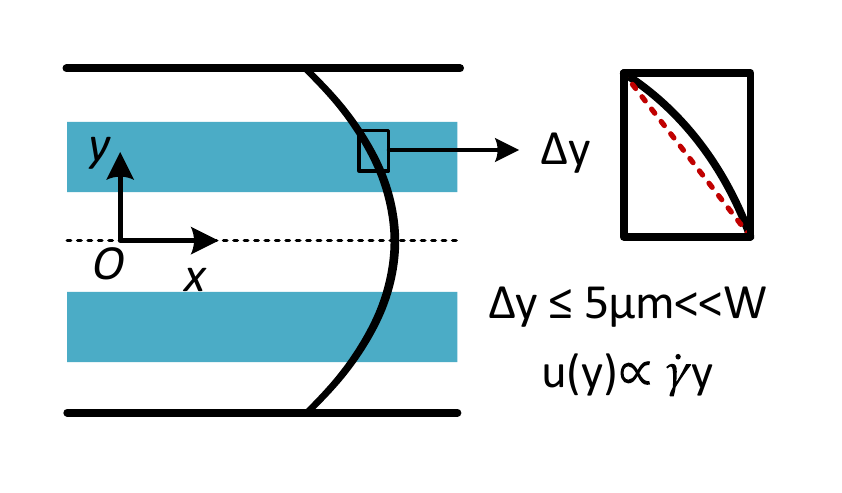}\\
	\vspace*{-0.3cm}
	\includegraphics[width=6.45cm]{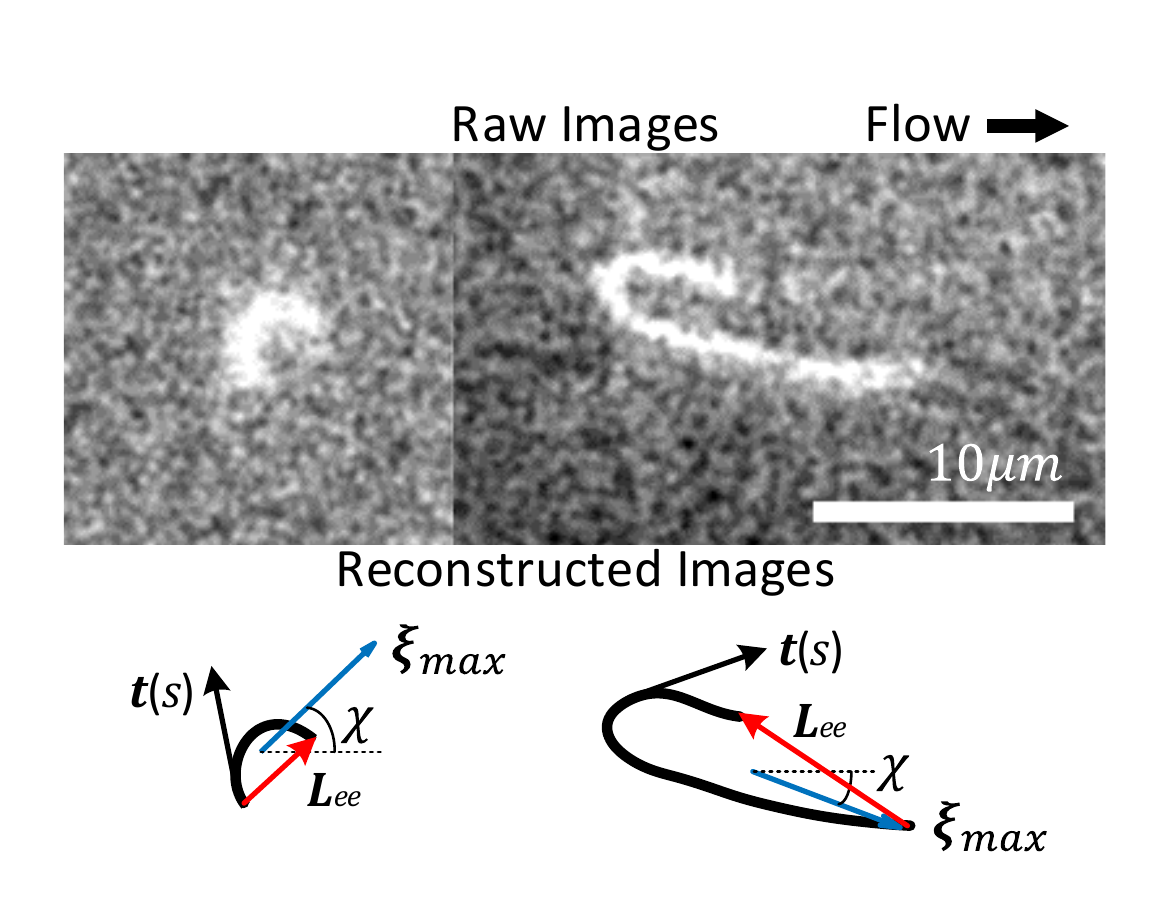}\vspace{-0.2cm}
	\caption{(\textit{A}) Sketch of the experimental setup. (\textit{B}) Velocity profile in the observation plane. Measurements take place in the blue regions, where the profile can be approximated as linear on the typical transverse scale $\Delta y$ of a filament. (\textit{C}) Raw images and reconstructed images of two configurations with corresponding parameters. }
	\label{Fig:setup}
\end{figure}

The microfluidic channel used in experiments is a vertical Hele-Shaw cell in which we drive a horizontal flow as illustrated in figure \ref{Fig:setup}\textit{A}. The channel dimensions are designed based on two factors: (1) The aspect ratio between height and width should be large enough to provide a velocity plateau in the $z$ direction near the channel center as soon as $z\gtrsim W$; in this region, the flow is nearly two-dimensional and the filaments are mainly deformed by the imposed shear rate $\dot{\gamma}_{y}$ in the $xOy$ plane; (2) The width of the channel is much larger than the typical transverse length scale $\Delta y$ of the deformed filaments, so that the shear can be approximated as constant over that length scale. To avoid wall-interaction effects and artefacts due to the change in sign of $\dot{\gamma}_y$ near the centerline, we focus on trajectories of filaments flowing in the blue region in figure \ref{Fig:setup}\textit{B}. Figure \ref{Fig:setup}\textit{C} shows raw images and reconstructed filaments for two different configurations as well as corresponding parameters: $\bm{L}_{ee}$ (red) is the end-to-end vector, $\bm{t}(s)$ (black) is the tangential vector, $\bm{\xi}_{max}$ (blue) is the eigenvector corresponding to the largest eigenvalue of the gyration tensor of the filament, and $\chi$ is the angle between $\bm{\xi}_{max}$ and the flow direction.

\section{Computational model and methods}

Actin filaments considered in this work have a characteristic diameter $d \sim 8\,$nm and typical lengths in the range of  $L \sim 4-40\, \mu$m. Due to the slenderness of these filaments (aspect ratio $\epsilon\equiv d/L\ll 1$), we opt to describe them as space curves parameterized by arc length $s\in[0,L]$, with Lagrangian marker $\bm{r}(s,t)$ denoting the position of any point along the centerline. Under over-damped conditions typical of microscale flows, seeking a balance between forces due to stretching and viscous drag provides the characteristic time scale for relaxation of a stretching mode as $\tau_s \sim \mu(L/d)^2/Y$, where $Y$ denotes Young's modulus \cite{powers2010dynamics}. A similar balance between bending forces and viscous drag provides the relaxation time of bending modes as $\tau_b \sim \mu(L/d)^4/Y$. The ratio of these two time scales is $\tau_s/\tau_b \sim (L/d)^{-2} \sim \OR(10^{-6})$, which implies that stretching modes relax much faster than bending modes. Consequently, we can approximate the filaments as inextensible, which results in a metric constraint on the Lagrangian marker: $\bm{r}_s \cdot \bm{r}_s  =1$, where indices denote partial differentiation and $\bm{r}_s=\hat{\boldsymbol{t}}$ is the local tangent vector. The bending energy of a filament with contour length $L$ is given by:
\begin{equation}
E = \frac{B}{2} \int_{0}^{L}{| {\bm{r}_{ss}} | }^2ds,
\label{Eq:bending E}
\end{equation}
where $B$ is the bending rigidity of the filament, which also defines its persistence length $\ell_p=B/k_B T$. The phalloidin-stabilized actin filaments considered here have a persistence length of $\ell_p=17\pm1\,\mu$m \cite{gittes1993flexural,isambert1995flexibility}.\ We model these filaments as inextensible fluctuating Euler-Bernoulli beams whose hydrodynamics in an imposed flow $\boldsymbol{U}_0(\bm{r})=\dot{\gamma}y$ are described by non-local slender-body theory \cite{tornberg2004simulating,Johnson1980} as
\begin{equation}
8\pi\mu(\bm{r}_t-\bm{U}_0(\bm{r}))=-\Lambda[\bm{f}](s)-K[\bm{f}](s), 	\label{Eq:non-localSBT1}
\end{equation}
with
\begin{align}
& \Lambda[\bm{f}] = [-c(\bm{I} + \bm{r}_s\bm{r}_s ) + 2(\bm{I} - \bm{r}_s\bm{r}_s )]\cdot \bm{f}(s)  \label{Eq:local operator},\\
& K[\bm{f}] = \int_{0}^{L} \left( \frac{\bm{I} + \hat{\bm{R}}  \hat{\bm{R}} }{|\hat{\bm{R}}|} \bm{f}(s') -  \frac{\bm{I} + \bm{r}_s \bm{r}_s }{|s-s'|}\bm{f}(s) \right) ds'.  \label{Eq:integral operator}
\end{align}
Here, $\mu$ is the suspending fluid viscosity, $\hat{\bm{R}}=\bm{r}(s)-\bm{r}(s')$, and $c=-\ln(e \epsilon^2)$ is a geometric parameter. $\Lambda$ is a local mobility operator that accounts for drag anisotropy, while the integral operator $K$ captures the effect of  hydrodynamic interactions between different parts of the filament. The force per unit length  $\bm{f}(s)$ has contributions from bending and tension forces as well as Brownian fluctuations: 
\begin{equation}
\bm{f}(s,t)=B\bm{r}_{ssss}-(\sigma(s)\bm{r}_s)_s+\bm{f}^{br},	\label{Eq:non-local SBT2}
\end{equation}
where $\sigma(s)$ is the Lagrange multiplier that enforces the constraint of inextensibility and can be interpreted as internal tension. The Brownian force density $\boldsymbol{f}^{br}$ obeys the fluctuation-dissipation theorem \cite{munk2006dynamics,manikantan2013subdiffusive}:
\begin{align}
\langle \bm{f}^{br}(s,t) \rangle&=\mathbf{0},\\
\langle \bm{f}^{br}(s,t)\bm{f}^{br}(s',t') \rangle&=2k_BT\Lambda^{-1}\delta(s-s')\delta(t-t').
\end{align}
Since the filament is freely suspended, we apply force- and moment-free boundary conditions at both filament ends, which translate to: $\bm{r}_{sss}=\bm{r}_{ss}=\sigma=0$. 

We non-dimensionalize the governing equations by scaling spatial variables with $L$, time by the characteristic relaxation time $8\pi\mu L^4/B$, the external flow by $L\dot{\gamma}$, deterministic forces by the bending force scale $B/L^2$  and Brownian fluctuations by $\smash{\sqrt{L/\ell_p}B/L^2}$ \cite{manikantan2013subdiffusive,munk2006dynamics}. The dimensionless equations are given as follows:  \vspace{-0.1cm}
\begin{align} 
& \bm{r}_t = \bar{\mu }\bm{U}_0(\bm{r}(s,t)) - \Lambda[\bm{f}](s)  - K[\bm{f}](s)  	\label{Eq:non-localSBTND1},\\
& \bm{f}=\bm{r}_{ssss}-(\sigma(s)\bm{r}_s)_s+ \sqrt{L/\ell_p}\,\boldsymbol{\zeta}	\label{Eq:non-localSBTND2}, \vspace{-0.1cm}
\end{align}
where two-dimensionless groups govern the dynamics: the elasto-viscous number $\bar{\mu}= 8\pi\mu\dot{\gamma} L^4/B$ is the ratio of the characteristic flow time scale to the time scale for elastic relaxation of a bending mode, while $L/\ell_p$ compares the filament contour length to its persistence length and measures the magnitude of thermal fluctuations. The random vector $\boldsymbol{\zeta}$ is uncorrelated in space and time and drawn from a Gaussian distribution with zero mean and unit variance.  Eqs.~(\ref{Eq:non-localSBTND1})--(\ref{Eq:non-localSBTND2}) are numerically integrated in time using an implicit-explicit time-stepping method that treats the stiff linear terms coming from bending elasticity implicitly and non-linear terms explicitly.  At every time step, the unknown tensions are obtained by solution of an auxillary dense linear system that can be derived from the intextensibility condition: $\bm{r}_{ts} \cdot \bm{r}_s = 0$. Further details of the numerical method can be found in \cite{tornberg2004simulating,manikantan2013subdiffusive}. All simulations presented here were carried out in two dimensions using $N=64$ points along the arc length of the filament. Typical time steps for the simulations were in the order of $\Delta t \sim 10^{-10}-10^{-12}$.


\section{Theoretical model}

\subsection*{Dynamics of the $J$-shaped configuration} 
\begin{figure}[t]
	\centering
	\hspace*{-7mm}
	\includegraphics[scale=0.4]{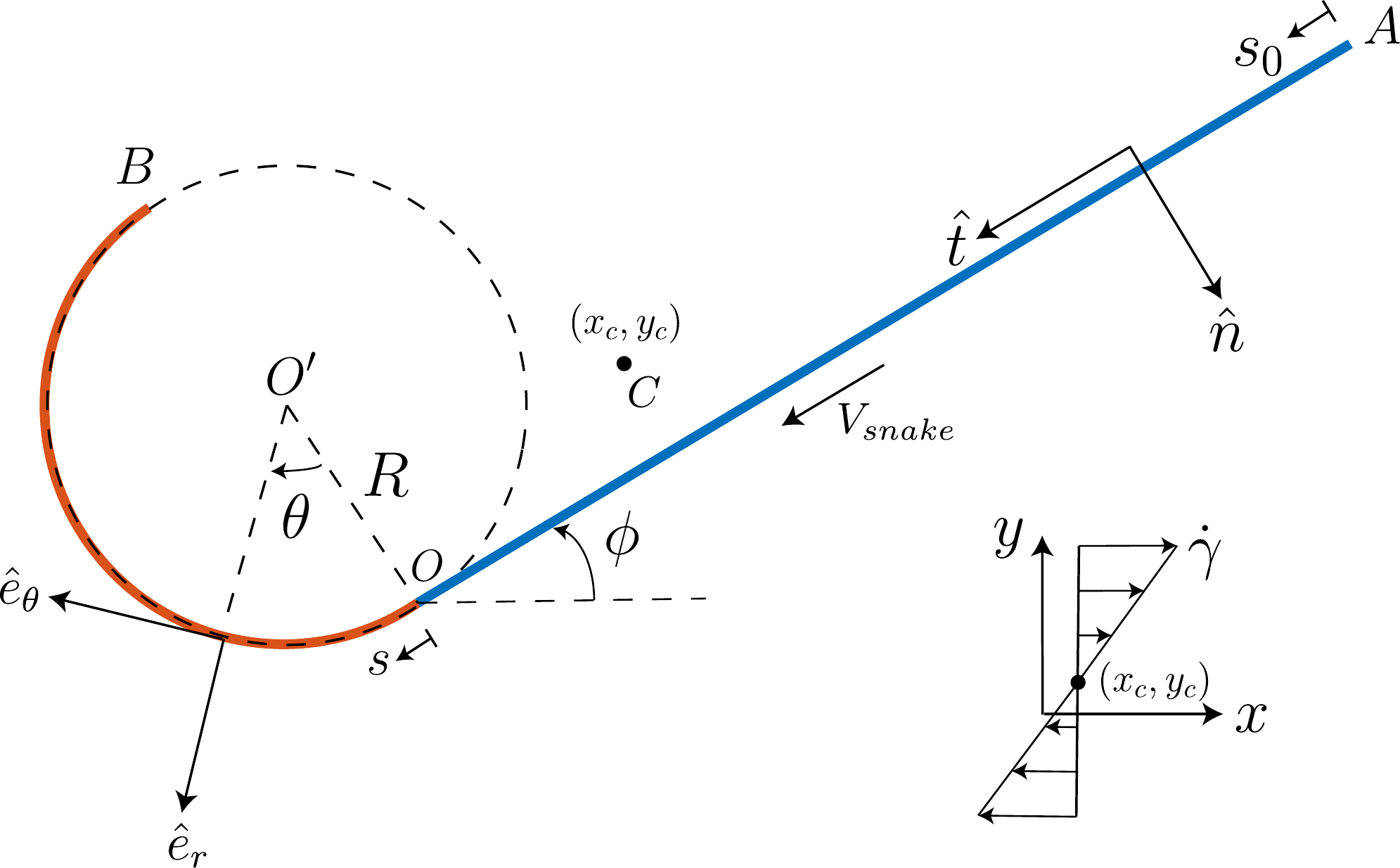}
	\caption{Theoretical model of the $J$ configuration. The bent part is approximated by a semi-circle of radius $R$. There is a snaking velocity $V_{snake}$ along the contour of the shape. The straight part of the configuration has a tilt angle of $\phi$ with the direction of the flow.}
	\label{Fig:tilted}
\end{figure}

The initiation of a $U$ turn in both experiments and simulations involves the formation of a $J$-shaped configuration which is tilted with respect to the flow direction and is a precursor to the snaking motion. To understand the transition to $U$ turns, we seek a simplified model of this configuration using the geometry shown in Fig.~\ref{Fig:tilted}. We approximate the bent portion of the filament by a semi-circle of yet unknown radius $R$ and assume that the rest of the filament is straight and has a tilt angle $\phi$ with respect to the flow direction.  We also introduce the following notations:

\begin{itemize}
	\setlength\itemsep{0.0em} 
	\item $T^{(1)} \equiv$ velocity of the straight arm in the tangential direction $\hat{\boldsymbol{t}}$.
	\item $v_\perp^{AO} \equiv$ velocity of the straight arm in the normal direction $\hat{\boldsymbol{n}}$.
	\item $T(\theta) \equiv$ velocity of the semi-circle along $\hat{\boldsymbol{e}}_\theta$.
	\item $v_\perp^{OB}(\theta) \equiv$ Velocity of the semi-circle in the $\hat{\boldsymbol{e}}_r$ direction.
	\item $C (x_c,y_c) \equiv$ filament center of mass.
	\item $(x,y) \equiv$ global coordinate axes centered at $O$.
	\item $l \equiv$ length of the straight arm, also given by: $l = L - \pi R$ where $L$ is the filament contour length.
\end{itemize}

Note that the assumption of a semi-circular shape for the bend leads to some inconsistencies. In particular, it is not possible to satisfy the force- and moment-free boundary conditions at point $B$. Adding a second straight arm emanating from $B$ would allow circumventing this issue, and the model we present here is justified in the limit of the length of that second arm becoming zero. Additional inconsistencies also arise at point $O$, where not all derivatives of the filament shape are continuous. These assumptions are necessary to make analytical progress, and we will see \textit{a posteriori} that the model produces results that are in good agreement with experimental and simulation data. As we discuss later, the model does also satisfy a global energy balance that serves to make the assumptions rigorous while neglecting the boundary layers that may arise at geometric discontinuities.

With the definitions above, the relative velocity between the fluid and the straight arm in the tangential and normal directions can be expressed as:
\begin{align}
v_\pll^{rel} &= T^{(1)} + \dot{\gamma}\left[ (l-s_0) \sin \phi - y_c\right] \cos \phi, \\
v_\perp^{rel} &= v_\perp^{AO}(s_0) - \dot{\gamma}\left[ (l-s_0) \sin \phi - y_c\right] \sin \phi.
\end{align}
As there are no forces acting in the normal direction inside the straight arm, we set $v_\perp^{rel}=0$ which yields
\begin{equation}
v_\perp^{AO}(s_0) = \dot{\gamma}\left[ (l-s_0) \sin \phi - y_c\right] \sin \phi.
\end{equation}
In the tangential direction, the internal tension $\sigma(s)$ induces an elastic force density $\boldsymbol{f}(s)=\sigma_s \hat{\boldsymbol{t}}$. This force density is balanced against viscous stresses using resistive force theory as $-\sigma_s=c_\parallel v_\pll^{rel}$, which can be integrated using the force-free boundary condition at point $A$ to yield
\begin{align}
\begin{split}
\sigma(s_0) =& -c_\pll T^{(1)} s_0 \\
& +c_\pll \dot{\gamma}\left[ \left(l s_0 - \tfrac{1}{2}s_0^2\right) \sin \phi  - y_c s_0\right] \cos \phi.
\end{split}
\end{align}
We have introduced the coefficient of resistance per unit length in the tangential direction, which is expressed as
\begin{equation}
c_\pll \approx \frac{2\pi\mu}{\log(2L/d)},
\end{equation}
and we similarly define $c_\perp\approx 2 c_\pll$ as the resistance coefficient for transverse motion. 

We analyze the kinematics and force balance on the semi-circular arc in a similar fashion and first express the relative velocities along the arc as
\begin{align}
v_\pll^{rel} &= T + \underbrace{v_f \cos (\theta-\phi)}_{v_\pll^f}, \label{eq:vrelpll} \\
v_\perp^{rel} &= v_\perp^{OB} + \underbrace{v_f \sin (\theta-\phi)}_{v_\perp^f},
\end{align}
where $v_f = \dot{\gamma} R \left[\cos \phi - \cos(\theta-\phi)\right] -\dot{\gamma}y_c$. Seeking a balance between elastic and viscous forces in the tangential and normal directions, we obtain:
\begin{align}
-\frac{1}{R} \frac{d \sigma}{d\theta} &= c_\pll v_\pll^{rel}, \\
\frac{\sigma}{R} + \frac{B}{R^3} &= c_\perp v_\perp^{rel}.
\end{align}
The constraint of inextensibility introduces a kinematic relation between the Lagrangian velocities in the tangential and perpendicular directions everywhere along $OB$, and provides the condition:
\begin{equation}
\frac{d T}{d\theta} + v_\perp^{OB} = 0. \label{eq:inext}
\end{equation}
Eqs.~(\ref{eq:vrelpll})--(\ref{eq:inext}) can be combined to yield a second-order non-homogeneous ODE for $v_\perp^{OB}(\theta)$:
\begin{equation}
2 \frac{d^2 v_\perp^{OB}}{d\theta^2} - v_\perp^{OB} = 2 \frac{d^2 v_\perp^f}{d\theta^2} + \frac{d^2 v_\pll^f}{d \theta^2}.
\end{equation}
This ODE can be solved analytically subject to continuity of the velocity at point $O$ and to the tension-free boundary condition at point $B$:
\begin{equation}
\begin{split}
v_\perp^{OB}(\theta) =&\,\, C_1 \cosh(\lambda \theta) + C_2 \sinh(\lambda \theta) \\
+& \sum_{n=1,2} [\alpha_n \cos(n \theta)+ \beta_n \sin(n \theta)],
\end{split}
\end{equation}
where $\lambda=1/\sqrt{2}$ and
\begin{align}
C_1&=\frac{\dot{\gamma}R\sin 2\phi}{18}, \\
C_2&=\frac{B}{c_\perp R^3 \sinh (\pi\lambda)}-\frac{\dot{\gamma}R}{18}\sin 2\phi \tanh \left(\frac{\pi \lambda}{2}\right), \\
\alpha_1&=\dot{\gamma}(y_c - R\cos\phi)\sin\phi, \quad \alpha_2=-\frac{5}{9}\dot{\gamma}R \sin 2\phi, \\
\beta_1&=-\dot{\gamma}(y_c - R\cos\phi)\cos\phi, \quad \beta_2=\frac{5}{9}\dot{\gamma}R \cos 2\phi.
\end{align}
From $v_\perp^{OB}$, the tangential velocity along the bend is easily inferred as
\begin{equation}\label{tanvel}
T = v_\pll^f - 2 \left(\frac{d v_\perp^{OB}}{d \theta} - \frac{d v_\perp^f}{d \theta} \right).
\end{equation}
Seeking continuity of tangential velocity and internal tension at point $O$, we obtain two distinct expressions for the tangential velocity $T^{(1)}$ of the straight arm:
\begin{align}
\begin{split}
T^{(1)} =& -\frac{2 B \lambda}{c_\perp R^3 \sinh(\pi \lambda)} + \dot{\gamma} y_c \cos \phi \\
&- \frac{2}{9}\dot{\gamma} R \cos 2\phi + \frac{\dot{\gamma}R \lambda}{9} \sin 2 \phi \tanh\left(\frac{\pi \lambda}{2}\right),  
\end{split} \label{cons1} \\
T^{(1)} =& \frac{B}{c_\pll R^2 l} - \frac{\dot{\gamma} l}{4}\sin 2 \phi + \dot{\gamma}y_c \cos \phi.\label{cons2}
\end{align}
Interestingly, \eqref{cons2} can be shown to also satisfy the torque balance on the filament.

For consistency, we require that Eqs.~(\ref{cons1})--(\ref{cons2}) be equal.\ Note, however, that both $R$ and $\phi$ remain unknown at this point.\ We therefore seek a third condition based on dissipation arguments similar to those used to explain the tank-treading motion of vesicles \cite{rioual2004analytical}. Over the course of an infinitesimal time interval $\delta t$ during a $U$ turn, a length of $\delta L \equiv V_{snake} \delta t$ that was initially straight becomes bent into the semi-circular curve of radius $R$, where $V_{snake}$ is the snaking velocity. During that same time, the same small amount of length becomes straight on the other side of the bend. The amount of work required to bend the straight part can be estimated as the change in its elastic energy:
\begin{equation}
\delta E = \frac{B}{2}\frac{V_{snake} \delta t}{R^2}. \label{eq:deltaE}
\end{equation}
This expression provides an estimate for the rate of change $\dot{E}=\delta E/\delta t$ of bending energy due to the deformation of the filament at it undergoes snaking. An alternative expression can be obtained from first principles by differentiating the bending energy as:
\begin{equation}
\dot{E} = B\int_0^L \br_{ss} \cdot \br_{tss}\, ds.
\end{equation}
Applying two integrations by parts and using the fact that tension forces do not perform any work leads to:
\begin{equation}\label{eq:work}
\dot{E} = B\int_0^L \br_{t} \cdot \br_{ssss} ds=\int_0^L \br_{t} \cdot \bff^e ds \equiv - \int_0^L \br_{t} \cdot \bff^h ds.
\end{equation}
where ${\boldsymbol{r}_t}$ is the velocity of a material point along the filament, and $\bff^e=B\br_{ssss}$ is the local elastic force density whose work balances that of the hydrodynamic force density $\boldsymbol{f}^h=-[c_\pll \boldsymbol{tt}+c_\perp (\boldsymbol{I}-\boldsymbol{tt})]\cdot\boldsymbol{v}^{rel}$. The detailed expression for the integral in Eq.~(\ref{eq:work}) is cumbersome and therefore omitted here. The above derivation involves integration by parts and assumes continuity of derivatives. Equating (\ref{eq:work}) with $\delta E/\delta t$ from Eq.~(\ref{eq:deltaE}), and identifying the snaking velocity with the tangential velocity $T^{(1)}$ of the straight arm in the $J$ shape, we obtain the additional condition:
\begin{equation}
\frac{BT^{(1)}}{2R^2}= -\int_0^L {\boldsymbol{r}_t}\cdot\boldsymbol{f}^h\,ds. \label{tank}
\end{equation} 
The above relation is essentially an integral energy balance in the system where we have included the dominant terms that come from the approximated $J$ shape. In principle, there may be other terms arising from boundary layers near the junctions of approximate straight and semi-circular arcs which are ignored here in an asymptotic sense to facilitate analytical progress.

It is possible to recombine \eqref{cons1}, \eqref{cons2} and \eqref{tank} to form two equations for the unknowns $R$ and $\phi$. These two equations are then solved numerically using a Newton-search algorithm.\ The equations essentially specify two curves in the $\phi-R$ plane, and a solution only exists when the curves intersect.\ For a given aspect ratio of the filament, we find that there exists a critical value of $\bar{\mu}$ below which the curves do not intersect. This suggests that below this  value   $J$-shapes can no longer form and therefore $U$ turns cannot occur. The theoretically calculated value of $\bar{\mu}_c^{(2)}/c\approx 1700$ is plotted as a dashed line in Fig.~4 of the main article and indeed provides a very good estimate for the onset of $U$ turns. 

\begin{figure}[t]
	\centering
	\includegraphics[scale=0.7]{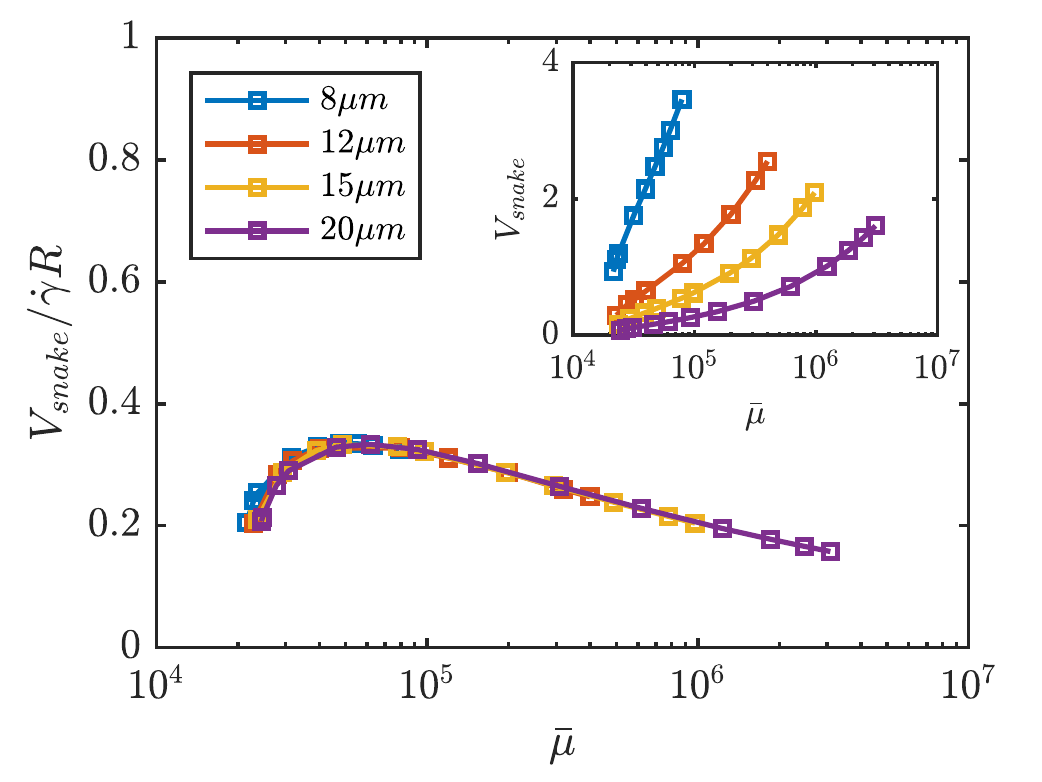} \vspace{-0.3cm}
	\caption{Snaking velocity $V_{snake}$ scaled by $\dot{\gamma}R$ as a function of  $\bar{\mu}$ as predicted by our theoretical model. Inset: same data before rescaling by $\dot{\gamma}R$. }
	\label{Fig:ScaledT}
\end{figure}

The model also provides the theoretical snaking velocity for the straight segment $OA$. The snaking velocities for different filaments with varying lengths are plotted in the inset of the Fig.~\ref{Fig:ScaledT} as a function of  elasto-viscous number. After rescaling with the product of the bend radius $R$ and shear rate $\dot{\gamma}$, all the data for different filament lengths collapse onto a master curve that only depends weakly on $\bar{\mu}$ in Fig.~\ref{Fig:ScaledT}. This collapse therefore confirms that the relevant dynamic length and time scales during the snaking motion are  $R$ and $\dot{\gamma}^{-1}$, respectively.

\section{Measurement of the bend radius}

\begin{figure}[H]
	\centering
	\hspace*{-7mm}
	\includegraphics[scale=0.85]{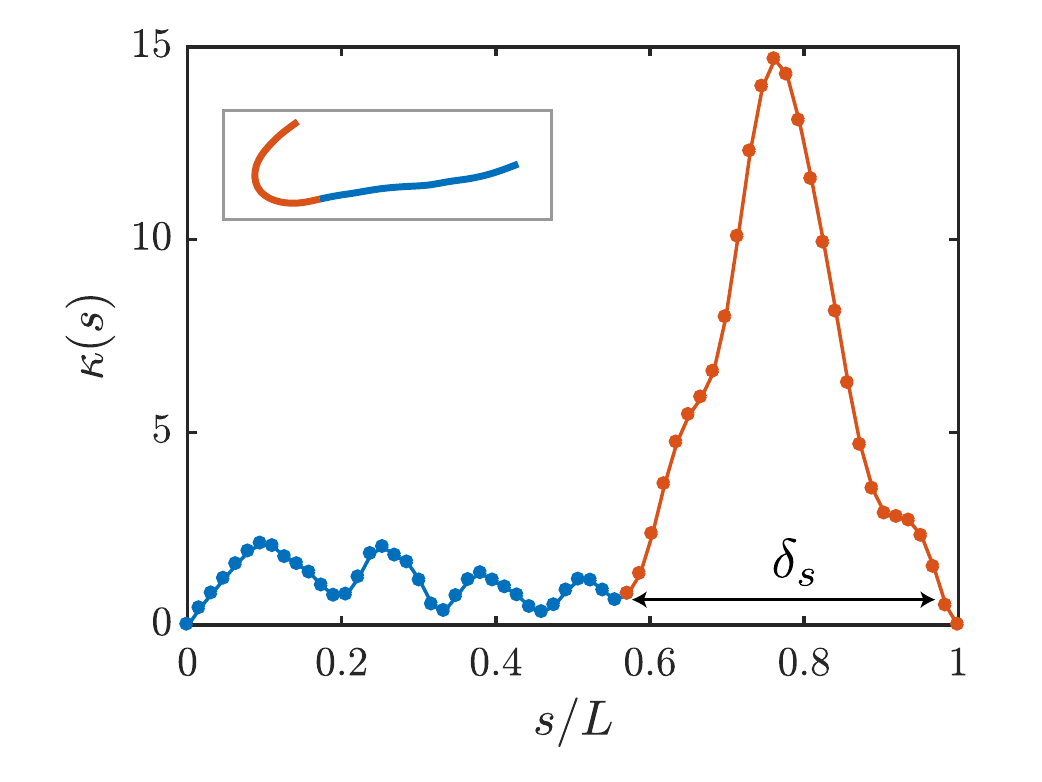}
	\vspace{-0.25cm}
	\caption{Variation of curvature $\kappa$ along the filament centerline for a typical $J$ shape chosen from a simulation. The length marked as $\delta_s$ provides an estimate for the arclength of the bent portion.}
	\label{Fig:kappa}
\end{figure}

Our theoretical model approximates the $J$ shape by a straight segment and a semi-circular arc. In this idealized configuration, the curvature is zero along the straight segment and then constant at $1/R$ over a length of $\delta_s=\pi R$. In experiments and simulations, however, the curvature varies smoothly and must reach zero at $s=L$ due to the boundary conditions. A typical curvature profile from a simulation is shown in Fig.~\ref{Fig:kappa}. In order to estimate the radius $R$ in a way that is consistent with the model, we measure the arclength $\delta_s$ over which the curvature, which increases from zero at $s=L$, decreases again to reach nearly zero. This measured length from simulations and experiments is shown in Fig.~6\textit{B} of the main article and is in good agreement with the predictions from our model.

An alternative measure of the radius can also be obtained from the plateau of the bending energy during a snaking turn as seen in Fig.~2\textit{C} of the main text. Since the majority of the energetic contribution comes from the sharp fold, we can get an estimate of the radius as
\begin{equation}
R=\frac{\delta_s}{\pi}\approx\frac{B \pi}{2 \langle E \rangle },
\end{equation}
where $\langle E \rangle$ is the average bending energy over the plateau. This measure is also plotted against the theoretical predictions in Fig.~6\textit{B} and follows similar trends.

In previous work, Harasim \textit{et al.}\ \cite{harasim2013direct} provided an expression for the bend radius that was independent of the length of the filament. For the parameter space explored in their study, they estimated $R \approx 1\, \mu$m. Our results partially agree with their finding in the limit of long filaments and strong shear. In our model, experiments and simulations, we find that the value of $\delta_s$ decreases weakly with flow strength and is in the range of $R \sim 0.7-1.5\, \mu$m.

\begin{equation}
R=\frac{\delta_s}{\pi}\approx\frac{B \pi}{2 \langle E \rangle },
\end{equation}

where $\langle E \rangle$ is the average bending energy over the plateau. This measure is also plotted against the theoretical predictions in Fig.~6\textbf{B} and follows similar trends.

In previous work, Harasim \textit{et al.}\ \cite{harasim2013direct} provided an expression for the bend radius that was independent of the length of the filament. For the parameter space explored in their study, they estimated $R \approx 1\, \mu$m. Our results partially agree with their finding in the limit of long filaments and strong shear. In our model, experiments and simulations, we find that the value of $\delta_s$ decreases weakly with flow strength and is in the range of $R \sim 0.7-1.5\, \mu$m.

\section{Onset of $J$ shape by global buckling}

While our theoretical model for the $J$ shape provides quantitative predictions for the onset of $U$ turns and parameters characterizing the shape and dynamics, the detailed mechanism for the formation of a $J$ shape from a nearly straight filament remains unclear. One mechanism, proposed by Lang et al. \cite{lang2014dynamics}, hypothesizes that the filament buckles locally over the characteristic length scale of transverse thermal fluctuations. However, the threshold derived from this local buckling hypothesis is inconsistent with the transition from $C$ buckling to $U$  turn found in our simulations and experiments.

\begin{figure}[H]
	\centering
	\hspace*{-1mm}
	\includegraphics[scale=0.8]{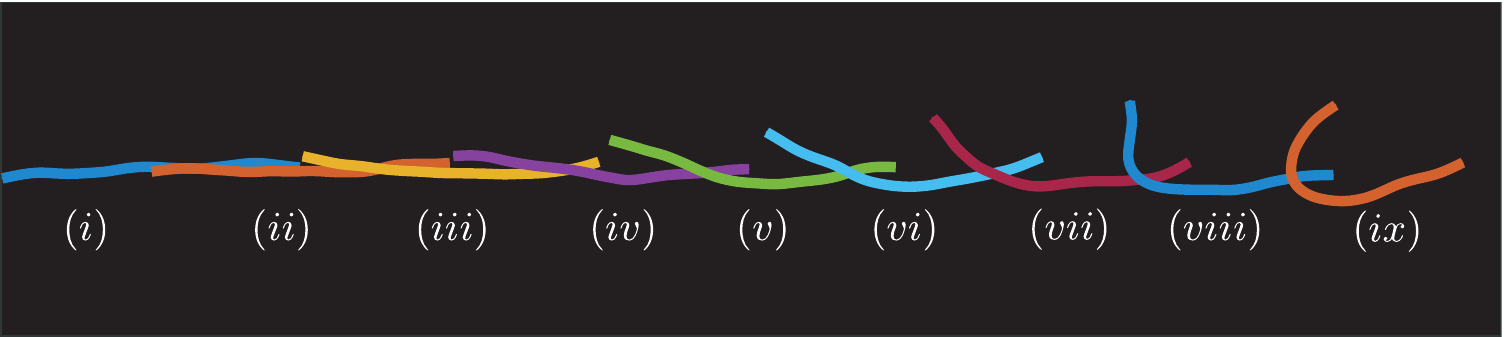}
	\caption{Numerical simulation showing the onset of a $J$ shape, which appears to result from buckling of the filament at a small angle.}
	\label{Fig:gb13}
\end{figure}

Another potential mechanism, which we elaborate on here, consists in global buckling of the filament at a small angle. To illustrate this mechanism, we show in Fig.~\ref{Fig:gb13} typical snapshots of filament configurations during the formation of the $J$ shape from a simulation. From these images, we see that the mean filament orientation enters the compressional quadrant of the flow before significant deformations arise (configurations iii and iv). As the filament starts to buckle under compressional viscous stresses (configurations iv and v), its changed shape causes portions of it to become aligned with the direction of extension, even though the mean orientation remains in the compressional quadrant. This results in a tension profile that changes sign along the filament, which in turn causes differential bending and migration of the high-curvature region from the filament center towards one of the ends (configurations vii and viii), thus giving rise to a $J$ shape (configuration ix). These findings are qualitatively different from $C$ buckling, where the entire filament experiences compression as it buckles during global rotation. 

\section{Characterization of the transition regime}

Brownian fluctuations are responsible for three effects: diffusion of the filament center of mass, rotational diffusion, and transverse shape fluctuations. In the present work, the effect of center-of-mass diffusion on the dynamics we observe can be neglected due to the large size of filaments. Orientational diffusion is well known to control the characteristic period of rotation in shear flow \cite{lang2014dynamics}, and the length scales of polymer extension and transverse thickness resulting from the balance between shear viscous forces and fluctuations are also directly linked to bulk shear viscosities \cite{Schroeder05}. Our results suggest, however, that Brownian fluctuations do not play a significant role in determining the onset of the buckling instability and the transition between $C$ and $U$ dynamics. Nonetheless, fluctuations tend to smooth the transitions between regimes, with transitional regions that become broader with decreasing $\ell_p/L$. 

\begin{figure}[H]
	\centering
	\hspace*{-7mm}
	\includegraphics[scale=0.75]{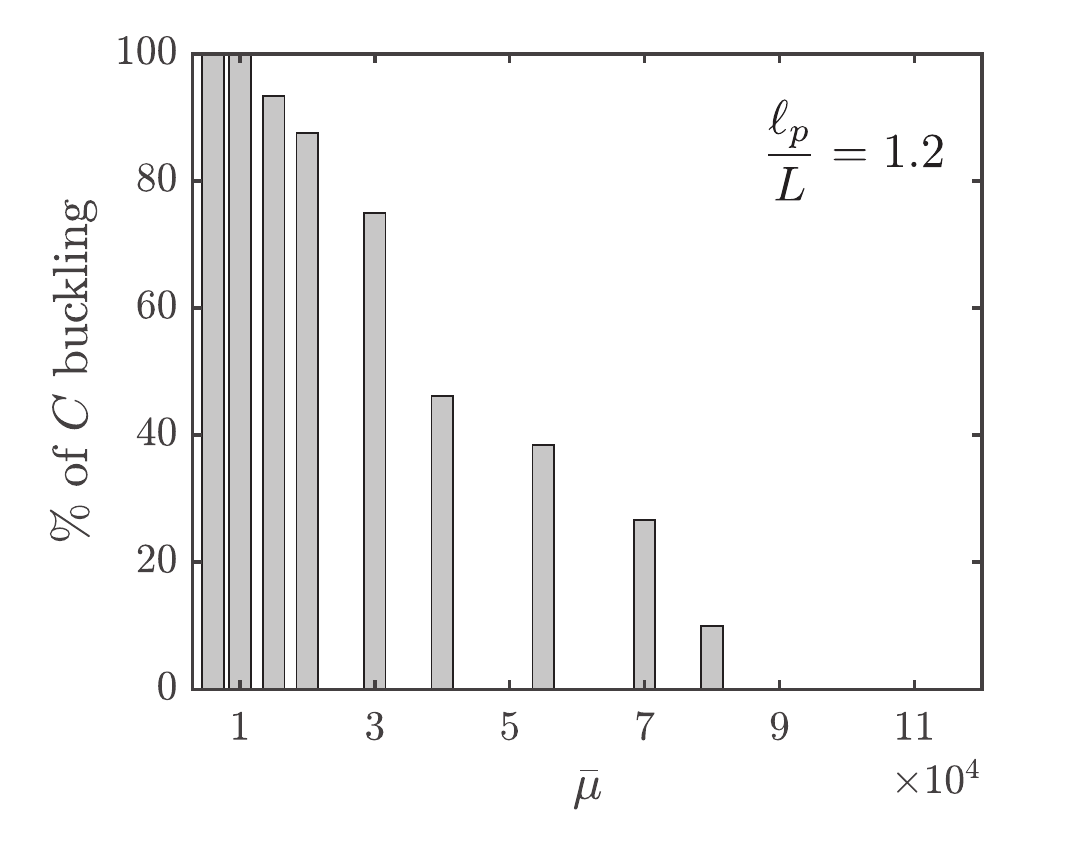}\vspace{-0.4cm}
	\caption{Percentage of $C$ buckling events as a function of $\bar{\mu}$ near the transition from $C$ buckling to $U$ turns in numerical simulations. The probability was estimated over a minimum of 10 distinct turns. The results shown are for $\ell_p/L = 1.2$.}
	\label{Fig:histogram1}
\end{figure}

Near the transition from $C$ buckling to $U$ turns, we have noted that both types of dynamics can occur over multiple tumblings of the end-to-end vector. This resulted in the gray area in Fig.~4 of the main article. This stochastic transitional regime can be characterized more precisely by the probability of observing either shape, which we can estimate in simulations and is shown in Fig.~\ref{Fig:histogram1} as a function of $\bar{\mu}$ for a fixed value of $\ell_p/L=1.2$. As expected, we find that the probability of $U$ turns continuously increases from 0 to 1 as $\bar{\mu}$ is varied across the transition. Similar stochastic transitions have been reported for the onset of buckling in compressional flows \cite{kantsler2012fluctuations,manikantan2015buckling}.

\vspace*{-0.3cm}

\section{Supplementary movie information}

\textbf{Movie 1 -- Jeffery tumbling experiments}: Movies from experiments are shown in \texttt{rod-experiments.mov}. The relevant parameters are ${\ell_p}/{L} = 3.75$ and $\bar{\mu} = 2.9 \times 10^3$.

\indent

\textbf{Movie 2 -- Jeffery tumbling simulations}: Movies from simulations are shown in \texttt{rod-simulation.mov}. The relevant parameters are same as above mentioned experiments.

\indent

\textbf{Movie 3 -- C buckling experiments}: Movies from experiments are shown in \texttt{C-experiments.mov}. The relevant parameters are ${\ell_p}/{L} = 2.9$ and $\bar{\mu} = 3.9 \times 10^3$.

\indent

\textbf{Movie 4 -- C buckling simulations}: Movies from simulations are shown in \texttt{C-simulation.mov}. The relevant parameters are same as above mentioned experiments.

\indent

\textbf{Movie 5 -- U turn experiments}: Movies from experiments are shown in \texttt{U-experiments.mov}. The relevant parameters are ${\ell_p}/{L} = 0.84$ and $\bar{\mu} = 2.1 \times 10^6$. 

\indent

\textbf{Movie 6 -- U turn simulations}: Movies from simulations  are shown in \texttt{U-simulation.mov}. The relevant parameters are same as above mentioned experiments. Over the multiple tumbling events shown in these movies we also observe an occasional $S$ turn.

\end{document}